# Low-Noble-Metal-Loading Hybrid Catalytic System for Oxygen Reduction Utilizing Reduced-Graphene-Oxide-Supported-Platinum Aligned with Carbon-Nanotube-Supported Iridium

Beata Dembinska[1,*], Magdalena Modzelewska[1], Agnieszka Zlotorowicz[1], Krzysztof Miecznikowski[1], Leszek Stobinski[2], Artur Malolepszy[2], Maciej Krzywiecki[3], Jerzy Żak[1,4], Enrico Negro[5], Vito Di Noto[5], Pawel J. Kulesza[1,*]

[1]*Faculty of Chemistry, University of Warsaw, Pasteura 1, PL-02-093 Warsaw, Poland*
[2]*Faculty of Chemical and Process Engineering, Warsaw University of Technology, Warynskiego 1, Warsaw, Poland*
[3]*Institute of Physics–CSE, Silesian University of Technology, Konarskiego 22B, 44-100 Gliwice, Poland*
[4]*Silesian University of Technology, Faculty of Chemistry, Strzody 9, 44-100 Gliwice, Poland*
[5]*Department of Industrial Engineering, Università degli Studi di Padova in Department of Chemical Sciences, Via Marzolo 1, 35131 Padova (PD) Italy*

\*Corresponding Authors

E-mail: *bbaranowska@chem.uw.edu.pl*; Tel.: +48225526336

E-mail: *pkulesza@chem.uw.edu.pl*; Tel.: +48225526344

**Abstract**

Hybrid systems composed of the reduced graphene oxide-supported platinum and multiwall carbon nanotubes-supported iridium (both noble metals utilized at low loadings on the level of 15 and $\leq 5$ µg cm$^{-2}$, respectively) have been considered as catalytic materials for the reduction of oxygen in acid media (0.5 mol dm$^{-3}$ H$_2$SO$_4$). The electrocatalytic activity toward reduction of oxygen and formation of hydrogen peroxide intermediate have been tested using rotating ring-disk electrode voltammetric experiments. The efficiency of the proposed catalytic systems has also been addressed by performing galvanodynamic measurements with gas diffusion electrode half-cell at 80 ºC. The role of carbon nanotubes is to improve charge distribution at the electrocatalytic interface and facilitate the transport of oxygen and electrolyte in the catalytic systems by lowering the extent of reduced graphene oxide restacking during solvent evaporation. The diagnostic electrochemical experiments reveal that at iridium-containing systems not only higher disk currents, but also much smaller ring currents have been produced (compared to reduced graphene oxide-supported platinum and its composite with bare carbon nanotubes), clearly implying formation of lower amounts of the undesirable hydrogen peroxide intermediate. The enhancement effect coming from the addition of traces of iridium (supported onto carbon nanotubes) to Pt, utilized at low loading, may originate from the high ability of Ir to induce decomposition of the undesirable hydrogen peroxide intermediate. There is a competition between activation (due to the presence of small amounts of Ir) and dilution (by carbon nanotubes) of Pt active centers in hybrid systems, therefore special attention is paid to the adjustment of their composition.

1. **Introduction**

Despite appreciable progress in the development of proton exchange membrane fuel cells, their commercialization still requires the lowering of the costs and more durable materials. The problem of electrochemical stability is crucially important for the cathode operating under oxidative conditions which include low pH, high potentials (especially during start-up/shut-down operations) and the generation of reactive oxygen species from hydrogen peroxide (the intermediate product of oxygen reduction) [1]. The latter issue may become even more serious when lower amount of expensive Pt catalyst is utilized, which on the one hand enables to reduce the costs of devices but on the other hand results in production of higher quantity of undesirable hydrogen peroxide (due to dilution of active centers). Further, high potential of the cathode may cause both the oxidative degradation of carbon support and the dissolution or sintering of platinum nanoparticles. Additionally, it is well recognized that platinum accelerates the corrosion rate of the carbon, what is particularly evident in a case of commonly used carbon blacks [2]. Therefore, there is need for searching alternative more corrosion-resistant carriers. In this regard, highly graphitized carbon materials like multiwalled carbon nanotubes (MWCNT) or more recently graphene-related materials are considered as promising alternatives for carbon blacks [3,4]. From the family of graphene-related materials the most perspective and the most widely used (due to facility of manufacturing and relatively low costs) are graphene oxide (GO) and reduced graphene oxide (rGO). The presence of surface oxygen-containing groups onto corrugated graphene layers facilitates their exfoliation and excellent dispersion of metal nanoparticles with narrow range of sizes what should allow for high utilization of the catalyst. But at the same time, large population of oxygen functionalities decrease the electronic conductivity of the support which, together with its restacking during slow evaporation of solvents, cause charge and mass (oxygen and water) transport limitations in catalytic layer [5]. One has to be aware that in the catalytic layer for oxygen electroreduction, where the reaction proceeds at three-phase boundary and under harsh conditions, the compromise in many planes should be assured. It has been proven for instance that for rGO-supported Pt the graphitization level and the content of oxygenated functionalities should be optimized not only in terms of proper electronic conductivity and stability of the rGO itself, but also to ensure the balance between hydrophobic and hydrophilic properties which in turn influence the dispersion of catalytic centers, the strength of their binding and the gas/water transport within catalytic material [6]. Further, several studies showed that the hybrid systems of two dimensional graphene-based materials with carbon blacks or carbon nanotubes (acting as separators between graphene sheets) allow the preparation of highly porous three dimensional structures with substantially improved mass transport and durability [7-15]. What is more, even an expansion of the d-spacing of the graphene layers in graphene nanosheets after assembling with carbon nanotubes or fullerene C60 has been observed [16]. Such composites, when applied as supports for platinum nanoparticles, have considerably increased the utilization of metallic centers during oxygen electroreduction process and the stability of the catalysts [11-15].

There have been many attempts of application of iridium and iridium(IV) oxide as the components of cathodic materials in polymer electrolyte membrane fuel cells (PEMFCs). Although their intrinsic activity is far away from the platinum-based catalysts, they have been effectively utilized as additives to platinum [17-36]. In this regard, depending on the route of coupling of Ir-based systems with Pt, different effects and mechanisms of their functioning were elucidated. In a case of metallic iridium additive, better efficiency of unsupported Pt-Ir alloys [18] or Pt-Ir nanoporous structures [19] than of Pt was attributed to the increase of the Pt 5d vacancy resulting in increased $O_2$ adsorption and weakening the O–O double bond. It was also observed that Ir(core)@Pt(shell) nanodendrites [20] and mixed platinum-iridium monolayers deposited on the surfaces of Pd(111) single crystals or carbon-supported Pd nanoparticles [21] owe their higher activity (compared to that of pure Pt counterparts) to lowering the coverage of Pt with OH as a result of lateral repulsion between the OH or O adsorbed on Ir and the OH adsorbed on a neighboring Pt atom. Similar activating effects (relative to Pt) were observed for PtIr dendritic tripods [22], PtIr sputtered thin films [23,24] and thin films of Pt deposited on Ir by thermal evaporation in vacuum [25]. It has also been reported more recently that PtIr alloy deposited by spray pyrolysis onto three dimensional crumpled rGO exhibited even higher activity than a commercially available Pt/C catalyst [26]. On the other hand, on the basis of the experiments conducted with carbon-supported core-shell structures (with Pt monolayer shell) and DFT calculations, it has been revealed that sole Ir core cause too strong contraction in Pt–Pt bonding and thus too weak binding energy of oxygen to Pt to dissociate it, therefore lower activity than Pt [27]. Also Popov et al. [28] reported decreased ORR efficiency of physical mixture of Pt/$TiO_2$ and Ir/$TiO_2$ in relation to pure Pt/$TiO_2$. It is worth mentioning, that none of described above research have discussed the possible changes in ORR mechanism on Pt due to its coupling with Ir. It has been shown that Ir nanostructures, despite of lower efficiency (in terms of less positive potentials), are more selective in the 4-electron ORR pathway with respect to the Pt [29,37]. It was observed that the addition of Ir to Pt promote the selectivity of the ORR in the 4-electron mechanism, which was ascribed to higher population of Pt active sites free of oxygen-based adsorbates [29].

Concerning $IrO_2$ utilization, there is general agreement in the literature that, regardless of the preparation method (physical mixture of Pt and $IrO_2$ or Pt nanoparticles deposited onto oxide support), the platinum activity decreases toward oxygen reduction in relation to unmodified Pt [30-36]. What is more, it has been shown that synthetized system of Pt/$IrO_2$ was less efficient than the physical mixture of Pt and $IrO_2$ blacks, what was ascribed to stronger adsorption of oxides on the platinum surface in the former case [31].

Finally, it is also well documented that the addition of Ir or $IrO_2$ to the cathodes of PEMFCs may considerably improve their durability [17,22,25,26,28,32-36,38], wherein the stability of $IrO_2$ was reported to be higher than metallic iridium [39]. It was proposed that these highly efficient oxygen evolution reaction catalysts suppress the electrochemical carbon corrosion due to decomposition of water around the carbon supports at high potentials [34-36].

In the present contribution, we explore the hybrid systems composed of the reduced graphene oxide-supported platinum and multiwall carbon nanotubes-supported iridium (both noble metals utilized at low loadings on the level of 15 and ≤ 5 μg cm$^{-2}$ under RRDE studies, respectively) as catalytic materials for the reduction of oxygen in acid medium (0.5 mol dm$^{-3}$ $H_2SO_4$). The role of carbon nanotubes is to improve charge distribution at the electrocatalytic interface and facilitate the transport of oxygen and electrolyte in the catalytic systems by lowering the extent of reduced graphene oxide restacking during solvent evaporation. Another important issue is the ability of the proposed carrier (carbon nanotubes decorated with Ir) to act as the efficient system of decomposition of the hydrogen peroxide intermediate, generated when Pt catalyst is utilized at low loading. The diagnostic electrochemical experiments reveal that at iridium-containing systems not only higher disk currents, but also much smaller ring currents have been produced (compared to reduced graphene oxide-supported platinum and its composite with bare carbon nanotubes), clearly implying formation of lower amounts of the undesirable hydrogen peroxide intermediate. Due to the competition between activation (by small amounts of Ir) and dilution (by carbon nanotubes) of Pt active centers in hybrid systems, special attention is paid to the adjustment of their composition.

## 2. Experimental

### 2.1. Chemicals

All chemicals were commercial materials of the highest available purity. 2-propanol, $HNO_3$, $H_2SO_4$, KOH were purchased from POCh (Poland), the solution of 5% Nafion-1100, platinum potassium hexachloroplatinate ($K_2PtCl_6$), multi-walled carbon nanotubes (O.D. $\times$ I.D. $\times$ L 10 nm $\times$ 4.5 nm $\times$ 3-6 μm) – from Sigma Aldrich, iridium(III) chloride hydrate – from Alfa Aesar. Nitrogen and oxygen gases (purity, 99.999%) were from Air Products (Poland). The solutions for electrochemical experiments were prepared from doubly-distilled and subsequently deionized (Millipore Milli-Q) water. Multi-walled carbon nanotubes (CNT) were activated in 2 mol dm$^{-3}$ $HNO_3$ under reflux for 2 h and washed with water until pH was near 6.

### 2.2. Materials preparation

Graphene oxide (GO) was prepared using a modified Hummer's method from commercial graphite powder (ACROS ORGANICS). In this process, 10 g of graphite powder was added to 230 ml of concentrated sulfuric acid (98 wt.%) and stirred for 30 minutes. Next 4.7 g of sodium nitrate and 30 g of potassium permanganate were slowly added to the mixture and the temperature was kept below 10 ºC in an ice bath. Then the mixture was slowly heated up to ~33 ºC and was controlled so as not to exceed 35 ºC for 2 hours under stirring. In the next step, 100 ml of water was added to the mixture and the temperature reached ~120 ºC. Finally, the mixture was treated with 10 ml of $H_2O_2$ (30 wt.%). Obtained slurry was kept in an ultrasonic bath for 1 hour. For purification, the slurry was filtered using ceramic membranes with 0.2 micron pore size and washed with deionized water in order to remove the by-products of the synthesis till the pH of the filtrate reached 6.5.

Reduced graphene oxide (rGO) was obtained in the course of the hydrazine reduction method in an analogous manner as described in ref [40]. In brief, 10 ml of 50 % hydrazine water solution was added to 100 ml of 0.5 wt.% GO water dispersion. The mixture was heated up to 100 °C and kept under stirring for 2 h. After reduction, the product was filtered using polyethersulfone (PES) filter with 0.8 μm pore size.

Reduced graphene oxide-supported platinum (Pt/rGO) catalyst (with final Pt loading at the level of 20% wt.) was synthetized in the course of the borohydride reduction method as described in ref [41]. Briefly, a proper amount of the Pt precursor ($K_2PtCl_6$) was added to 0.5 g rGO water dispersion under continuous magnetic stirring. After 15 min the 0.1 mol $dm^{-3}$ KOH was added to the mixture to increase the pH to 10. Nanoparticles of Pt were obtained by adding 10 ml of 1 mol $dm^{-3}$ $NaBH_4$ water solution to the slurry and kept under stirring. After 1 h the 20 wt.% Pt/rGO catalyst deposit was filtered using PES filter with 0.8 μm pore size and washed with deionized water.

The synthesis of carbon nanotubes-supported iridium nanoparticles (with final Ir loading at the level of 10% wt., Ir/CNT) was obtained through thermal reduction (under reflux for 2 h in $N_2$ atmosphere) of $IrCl_3$ in the solution of 2-propanol in the presence of carbon support. Before reflux, the suspension of CNT and solution of $IrCl_3$ were subjected to sonication for 1 h. After reflux the sample was washed out with water by centrifugation (3-4 times) and dried on a hot plate at 50 °C.

## 2.3. Equipment and characterization of materials

Electrochemical measurements were performed using CH Instruments (Austin, TX, USA) 920D workstation. A glassy carbon rod served as a counter electrode and the reference electrode was a mercury/mercurous sulfate electrode, $Hg/Hg_2SO_4$ (all potentials were recalculated and expressed versus the reversible hydrogen electrode (RHE) scale). The rotating ring-disk electrode (RRDE) voltammetric experiments were conducted via variable speed rotator (Pine Instruments, USA). RRDE assembly included a glassy carbon disk (diameter 5.61 mm) and a platinum ring (inner and outer diameters were 6.25 and 7.92 mm, respectively) and its collection efficiency (determined with the procedure reported previously [37]) was equal to 0.39. Prior the experiments the working electrode was polished with aqueous alumina slurries (grain size 0.05 μm) on a Buehler polishing cloth.

The inks for fabricating the electrocatalytic layers for thin-film RRDE studies (with constant platinum loading of 15 μg cm$^{-2}$) were prepared by sonicating and mixing at magnetic stirrer of 5 mg of catalysts (or hybrid systems), 1 ml of 2-propanol and 10 – 30 %ww of Nafion®. Due to different densities of carbon supports it was necessary to optimize the content of Nafion in each case (in terms of the highest possible homogeneity of the layers and obtaining proper ionic conductivity and mass transport). Therefore, for rGO-containing samples its optimum content was 10%, for CNT – 30% and for the mixtures it proportionally depended on the ratio of rGO and CNT. The inks were introduced on the GC electrode surface and, subsequently, dried at room temperature, 22±2 °C. The electrodes, covered with the catalytic layers, were washed out with the stream of water and subjected to potential cycling in the range between 0.05 V and 0.8 V (vs RHE) at the 100 mV s$^{-1}$ in the N$_2$-saturated 0.1 mol dm$^{-3}$ 0.5 mol dm$^{-3}$ H$_2$SO$_4$ until a stable stationary voltammetric responses were obtained. Before electrochemical examination, the hybrid systems were produced by subjecting Pt/rGO to mixing (sonicated and mixed at magnetic stirrer) with Ir/CNT or CNT. The mass ratio of Pt/rGO to Ir/CNT was 2.5 : 1 , 2 : 1 and 1.5 : 1. For comparative studies, a sample of Pt/rGO mixed in the mass ratio of 2 : 0.9 with CNT was prepared (the amount of CNTs was the same as in the sample Pt/rGO+Ir/CNT 2 : 1).

The performance of examined catalysts was also tested at 80 °C with a gas diffusion electrode (GDE, geometric area of active part, 3 cm$^2$) mounted into Teflon Flex Cell (Gaskatel GmbH, Germany). A spiral PtIr wire and Hg/Hg$_2$SO$_4$ electrode (the latter placed in a salt bridge) served as counter and reference electrodes, respectively. The gas diffusion backing layer for electrocatalysts was a carbon cloth with microporous layer (W1S1005, Fuel Cell Store, TX, U.S.A.). Catalytic layers were brush painted from the inks (using vacuum table heated up to 80 ºC), in which the ratio of catalyst to Nafion® was the same as in the course of RRDE thin film studies, but the level of dilution was higher (1 mg catalyst per 1 ml of 2-propanol). Final loading of Pt was about 50 μg cm$^{-2}$ and the loading of Ir was about 12.5 μg cm$^{-2}$. The current densities were normalized to the exact mass of platinum (determined by weighing the carbon cloth).

Transmission Electron Microscopy (TEM) experiments were carried out with Libra 120 EFTEM (Carl Zeiss) operating at 120 kV. Scanning electron microscopic (SEM) measurements and energy-dispersive X-ray analysis were achieved using MERLIN FE-SEM (Carl Zeiss) equipped with EDX analyzer (Bruker). The X-Ray powder diffraction (XRD) spectrum of Pt/rGO was measured using Rigaku-Denki X-ray diffractometer in a Bragg-Brentano configuration with a Ni filtered Cu radiation from a sealed tube operating at 40 kV and 40 mA stability of 0.01%/8 h. The measurement was carried out in the range of 20 to 125 degree with step 0.04 degree and counting time 10 s step$^{-1}$. The XRD pattern of Ir/CNT was collected using Bruker D8 Discover equipped with a transmission geometry and Cu x-ray source (l=1.540598 Å). The measurement was performed at scan rate 1.08 degree/min with 0.012 degree steps within a range from 10 to 130 degree. The data were analyzed using EVA software (Bruker AXS). XPS investigations were carried out in multi-chamber ultra-high vacuum experimental setup (base pressure 5 × 10$^{-8}$ Pa) equipped with PREVAC EA15 hemispherical electron energy analyzer with 2D-MCP detector. The samples were irradiated with an provided by an Al-K$_α$ X-ray source (PREVAC dual-anode XR-40B source, energy 1486.60 eV). For the survey spectra the pass energy was set to 200 eV (with scanning step 0.9 eV). Particular energy regions to 100 eV (with scanning step 0.05 eV). The binding energy (BE) scale of the analyzer was calibrated to Au 4f$_{7/2}$ (84.0 eV) [42]. Recorded data were fitted utilizing CASA XPS® embedded algorithms and relative sensitivity factors [43]. For background subtraction the Shirley function was used. If not specified, the components were fitted with a sum of Gaussian (70 %) and Lorentzian (30 %) lines. The full width at half maximum (FWHM) values for the peaks at the same binding energy region were allowed to vary within a narrow range. The estimated uncertainty for components' energy position determination was 0.1 eV.

## 3. Results and discussion

The Pt/rGO, Ir/CNT and the hybrid system of Pt/rGO+Ir/CNT (with 2 : 1 mass ratio) were examined using transmission and scanning electron microscopies. It is apparent from Fig. 1A that the diameters of platinum nanoparticles are on the level of 7-8 nm, but except the areas with homogenous distribution of "single" particles, there are also regions where their agglomerates can be found (inset to Fig. 1A and Fig. 1B). Fig. 1C illustrates TEM micrograph of iridium nanoparticles generated onto CNT support. Among important issues is the uniform arrangement of the Ir nanoparticles of low sizes (3-4 nm) and very low degree of agglomeration (which is also evident in SEM image presented in Fig. 1D where no big clusters of nanoparticles could be observed). Figs 1E and F show well-mixed nature of the hybrid system (Pt/rGO+Ir/CNT) indicating that mesoporous structure of CNT may create interfacial voids for better water and gas management and act as spacer preventing the restacking of graphene sheets. EDAX spectrum of the Ir/CNT gave the following chemical compositions (in ww %): Ir: 4.7%, C 95.3%. EDAX spectrum of the Pt/rGO gave the following chemical compositions (in ww %): Pt: 6.5%, C 93.5%.

Fig. 2 shows the results of XRD experiments conducted for the Pt/rGO and Ir/CNT samples. The XRD pattern of the Pt/rGO catalyst (Fig. 2A) exhibits a broad peak at 24.5 ° corresponding to (111) reflection of carbon and two small peaks at 42.3 ° (100) and 78.1 °

(110) indicating a short range order in stacked graphene layers [40]. The pattern contains also face centered cubic Pt peaks: (111), (200), (220), (311), (222), (400), (331) and (420) at 2θ values of 39.8 °, 46.3 °, 67.5 °, 81.3 °, 85.8 °, 103.6°, 117.8° and 122.9 °, respectively. The average crystallite size, determined from the peak (220) using Scherrer equation, is 7 nm and is consistent with the results form TEM experiments.

The XRD pattern of Ir/CNT (Fig. 2B) presents diffraction peaks near 26° (002), 44° (100), 54° (004) and 78° (110) characteristic of hexagonal structure of graphite in carbon nanotubes [41]. The peaks characteristic of face centered cubic crystalline iridium at about 41, 47, 69 and 84 should be ascribed to the (111), (200), (220) and (311) planes, respectively [37]. Low intensity of Ir signals and high full width at half maximum (fwhm) suggests very low size of crystallites and their low crystallinity. Therefore the average crystallite size – 3.8 nm – estimated from Pawleys' fitting and Rietved analysis should be treated with caution, however the data quite well agree with TEM microscopy (Fig. 1C).

To get insight into the chemical nature of iridium species in Ir/CNT we have performed the XPS measurement. Figs 3A and B present the XPS regions of Ir -4f and O 1s respectively recorded for Ir/CNT sample. Decomposition of the Ir- 4f region reveals two doublets resulting from spin-orbit splitting of 4f XPS line. The peaks of the first doublet (located at 61.3 and 64.5 eV) correspond to $4f_{7/2}$ and $4f_{5/2}$ lines of metallic form of the iridium. This particular component was fitted with Gaussian/Lorentzian product formula modified by the exponential blend. The binding energies of the second doublet (62.9 and 66.1 eV) should be ascribed to Ir-O bond [25,38]. The peak positions for both: metallic and oxidized iridium are consistent with other literature findings [44]. The relative intensity between metallic Ir and oxidized Ir components ($Ir_{metal}/Ir_{oxide}$ = 2.51) points that only a little bit more than 1/3 of the overall iridium amount was oxidized. This is highly probable, since the iridium creates agglomerates which are oxidized only at their surface leaving the core area unaltered in its metallic form.

The iridium is known from its high affinity to oxygen species, therefore the results are consistent with the view that small nanoparticles (as evident form TEM and XRD analyses) are prone to surface oxidation under exposure to air. As a confirmation to above statement the inset to Fig. 3A presents the Ir 4f region for control sample containing 2% of Ir (prepared in the same manner as the one with 10% of Ir). The decomposition revealed, according to peak binding energy position, that the whole amount of the iridium is present in its oxidized form [44]. Since from additional TEM experiments (data not shown) we could observe that for the 2% sample the Ir agglomerates were of significantly smaller dimensions, we assume that the whole volume of iridium was penetrated by oxidizing agents.

Analysis of O 1s spectrum (Fig. 3B) reveals three components at the binding energies of 531, 532.1 eV and 533.6 eV out of which two first shall be attributed to O-Ir and C-O bonds existing on the carbon support [45]. The third component is most likely originating from surface functionalities (like –COOH) present on oxidized CNT surface. The inset to Fig. 3B presents the O 1s region of control 2% sample. The relation in intensities between C-O components and O-Ir components for 2% and 10% samples supports previous statement on iridium oxidation form.

Fig. 4A shows cyclic voltammetric response of Ir/CNT layer deposited on glassy carbon electrode (solid line) in the deaerated 0.5 mol dm$^{-3}$ H$_2$SO$_4$ solution (the dotted line in Fig. 4A stands for response of unmodified CNT). The peaks observed at potentials below 0.35 V, characteristic to adsorption/desorption of hydrogen atoms at the metal surface, and surface oxide/hydroxide formation (starting just above the hydrogen region) are typical for Ir nanostructures [37,46-49].

Cyclic voltammograms of Pt/rGO (a) and hybrid systems obtained by mixing of Pt/rGO with various amounts of Ir/CNT (b-d) are presented in Fig. 4B. As described in the Experimental section, the mass ratio of Pt/rGO to Ir/CNT was correspondingly 2.5 : 1 (b) , 2 : 1 (c) and 1.5 : 1 (d). It is apparent form Fig. 4 that, except double layer charging currents increase, the rise of both hydrogen adsorption/desorption peaks as well as currents related to the formation of surface oxides are observed upon introduction of growing amounts of the Ir/CNT additive. What is more, when the responses of Pt/rGO+Ir/CNT(2:1) (curve c) are compared to Pt/rGO+CNT(2:0.9) sample (curve e), where the amount of CNT is maintained at the same level, high-order growth of the currents, corresponding to both noble metals electroactivity (Pt and Ir) and double layer charging, are evident in the former case. The results imply the more accessible electrochemically active surface area due to improved charge distribution and better hydrophilic properties of the layer following incorporation of Ir nanostructures.

Fig. 5 illustrates representative voltammetric disk (A) and simultaneous ring steady-state (B) currents recorded during the reduction of oxygen using bare Pt/rGO (a) and the hybrid catalysts composed of Pt/rGO admixed with Ir/CNT at 2.5:1 (b), 2:1 (c) and 1.5:1 (d) ratios. For comparative purposes also the results for the Ir-free system, namely Pt/rGO admixed with CNT at 2:0.9 ratio (curve e), are presented in Fig. 5. It is evident that the disk current densities have increased in a case of the system containing carbon nanotubes Pt/rGO+CNT (2:0.9) (curve e), when compared to the bare Pt/rGO catalyst (curve a). This may imply that the CNT not only improve charge distribution at the electrocatalytic interface, but also facilitate the transport of oxygen and electrolyte in the layer by lowering the extent of rGO restacking during slow solvent evaporation. Meanwhile, the ring currents have also increased to some extent (especially at less positive potentials), therefore it is evident that the presence of CNT slightly alters the mechanism of the process, probably due to "dilution" of Pt catalytic centers. On the other hand, at iridium-containing systems (curves b-d) not only higher disk currents (Fig. 5A), but also much smaller ring currents (Fig. 5B) have been produced (compared to bare Pt/rGO and Pt/rGO+CNT) clearly implying formation of lower amounts of the undesirable H$_2$O$_2$ intermediate. Finally, it is apparent from Fig. 5A that the current densities of oxygen reduction reach maximum values for the hybrid system of Pt/rGO+Ir/CNT(2:1) (curve c) and further increase of Ir/CNT additive (like in a case of Pt/rGO+Ir/CNT(1.5:1), curve d) does not improve the current densities. It is reasonable to expect that there is a competition between activation and dilution of Pt active centers in hybrid systems, therefore their composition should be adjusted with caution.

To estimate the percentage of H$_2$O$_2$ ($X_{H2O2}$) produced during oxygen reduction at catalytic films and number of electrons ($n_e$) participating in the electrocatalytic reaction, we have used the equations given below [50-52]:

$$X_{H_2O_2} = \frac{200 I_R / N}{I_D + I_R / N} \quad (1)$$

$$n_e = \frac{4 I_D}{I_D + I_R / N} \quad (2)$$

In the equations (1) and (2) $I_R$ is ring current, $I_D$ stands for disk current and $N$ is the collection efficiency of the RRDE assembly. It becomes apparent from Figs 6A and B, where $X_{H2O2}$ and $n_e$, both plotted versus potential applied to the disk electrode are presented, that the selectivity of the process toward 4-electron reduction (with formation of water as the major product) is increased in the presence of Ir additive.

When compared to platinum-based materials, iridium nanostructures are considered as relatively weak catalyst toward oxygen reduction [29,37], while their high activity in the electrochemical/chemical decomposition of hydrogen peroxide has been postulated [37,53].

Thus, in the next step we have evaluated the Ir/CNT as a potential catalyst of both processes, maintaining the same low loading as in the hybrid system of Pt/rGO+Ir/CNT(2:1) which could be viewed as the most active during reduction of oxygen (because it yielded the highest current densities). The RRDE voltammetric experiments conducted in $O_2$-staurated electrolyte (under the same conditions as those presented in Fig. 5) reveal that Ir/CNT reduces oxygen at much more negative potentials when compared to Pt-containing systems (starting and half-wave potentials are shifted to lower values of around 0.2 and 0.3 V, respectively, please compare Fig. 7A and Fig. 5A). What is more, as evident from Fig. 7B, observed ring currents are small indicating high selectivity of the compound toward production of water. The number of exchanged electrons (3.95) is at the same level as for Pt-containing hybrid systems (please compare Figs 7C and 6B), in spite of the fact that the concentration of active Ir sites is much lower (3.75 µg cm$^{-2}$) than Pt (15 µg cm$^{-2}$). Further experiment, conducted in deoxygenated 0.5 mol dm$^{-3}$ $H_2SO_4$ with addition of 1 mmol dm$^{-3}$ of $H_2O_2$ (i.e. with the concentration similar to the concentration of oxygen in its saturated electrolyte) revealed that $H_2O_2$ reduction starts at potential as high as 0.9 V, that is at least 0.1 V higher than for $O_2$ reduction, thus implying better efficiency of Ir nanostructures in the former process. The above observations imply that the enhancement effect coming from the addition of traces of iridium (supported onto CNT) to Pt/rGO (with low Pt loading) may originate from the high ability of Ir to induce decomposition of the undesirable hydrogen peroxide intermediate.

On the basis of our preliminary accelerated stability tests (1500 cycles at 50 mV s$^{-1}$ between 0.6 -1 V vs RHE, data not shown) we can conclude that the presence of carbon nanotubes increases the stability of Pt/rGO, meanwhile iridium gradually dissolves into the bulk of the solution (confirmed by EDX analysis).

Finally, we also performed galvanodynamic measurements with use of gas diffusion electrode mimicking the conditions existing in the low-temperature fuel cell. Fig. 8 shows typical polarization curves (potential vs. applied current density normalized to Pt mass, i.e. specific current) recorded in the oxygen saturated 0.5 mol dm$^{-3}$ $H_2SO_4$ at 80 °C following deposition of Pt/rGO (curves a and b) and hybrid system Pt/rGO+Ir/CNT(2:1) (curve c). In a case of Pt/rGO, although the electrode was cycled first in $N_2$-saturated and next in $O_2$-saturated electrolyte until stable responses were obtained, the subsequent first four polarization curves revealed gradual activation of the layer at higher specific currents (curves a and b). Such phenomenon was not observed for Pt/rGO+Ir/CNT(2:1) (curve c). Further, it is evident from Fig. 8 that, upon application of the same specific currents, in the region of activation polarization similar half-cell potentials are produced and even a slightly more favorable run for Pt/rGO can be observed. Meanwhile, with the increase of specific current, up to the region where gas transport should dominate, definitely higher half-cell potentials are produced at the hybrid Pt/rGO+Ir/CNT(2:1) system. The data of Fig. 8 are in almost perfect agreement with the results of diagnostic RDE experiments (Fig. 5A).

**Conclusions**

We demonstrate here that the addition of carbon nanotubes decorated with iridium nanoparticles results in the enhancement of electrocatalytic activity of graphene oxide-supported platinum during the reduction of oxygen in acid medium. The effect is demonstrated in terms of the higher catalytic current densities observed and much lower amounts of hydrogen peroxide produced under conditions of rotating ring disk voltammetric studies. The enhancement effect coming from the addition of traces of iridium (supported onto CNT) to Pt/rGO (with low Pt loading) may originate from the high ability of Ir to induce decomposition of the undesirable hydrogen peroxide intermediate. The improved efficiency of the hybrid system was confirmed by galvanodynamic measurements conducted at elevated temperature (80 ºC) with use of gas diffusion electrode operating under conditions resembling those characteristic of the cathode operating in a real fuel cell.

**Acknowledgments**

The support from the European Commission through the Graphene Flagship – Core 1 project [Grant number GA-696656] is highly appreciated. Authors acknowledge the ESPEFUM laboratory at the Institute of Physics – CSE, Silesian University of Technology for access to the UPS experimental setup.

**Figure captions**

**Fig. 1.** TEM and SEM images of (**A,B**) Pt/rGO,(**C,D**), Ir/CNT and (**E,F**) Pt/rGO+Ir/CNT.

**Fig. 2.** XRD patterns of Pt/rGO(**A**) and Ir/CNT (**B**).

**Fig. 3.** XP spectra of the Ir 4f (**A**) and O 1s (**B**) regions recorded for the Ir/CNT. The insets to panels (**A**) and (**B**) present the Ir 4f and O 1s regions recorded for 2% Ir control sample.

**Fig. 4.** Cyclic voltammetric responses recorded for catalytic films (deposited on glassy carbon disk of RRDE assembly). Electrolyte: argon-saturated 0.5 mol dm$^{-3}$ H$_2$SO$_4$. Scan rate: 10 mV s$^{-1}$.

**Fig. 5.** Normalized (background subtracted) rotating disk (**A**) and ring (**B**) voltammograms recorded during oxygen reduction at the catalytic films in oxygen-saturated 0.5 mol dm$^{-3}$ H$_2$SO$_4$ at the scan rate of 10 mV s$^{-1}$ and rotation rate of 1600 rpm.

**Fig. 6.** Formation of the hydrogen peroxide intermediate (**A**) and number of exchanged electrons (**B**) during oxygen reduction under conditions of RRDE voltammetric experiments in Fig. 5.

**Fig. 7.** Normalized rotating disk (**A**) and ring (**B**) voltammograms recorded during oxygen reduction at the Ir/CNT film in oxygen-saturated 0.5 mol dm$^{-3}$ H$_2$SO$_4$ at the scan rate of 10 mV s$^{-1}$ and rotation rate of 1600 rpm. (**C**) Normalized rotating disk voltammogram recorded at the Ir/CNT film in 0.5 mol dm$^{-3}$ H$_2$SO$_4$ in the presence of 1 mmol dm$^{-3}$ H$_2$O$_2$; scan rate: 10 mV s$^{-1}$; rotation rate: 1600 rpm. (**D**) The number of exchanged electrons during oxygen reduction at Ir/CNT under conditions of RRDE voltammetric experiment.

**Fig. 8.** Galvanodynamic steady-state polarization responses for the oxygen reduction on gas diffusion electrodes. Oxygen flux: 50 ml min$^{-1}$, temperature: 80 °C, electrolyte: oxygen saturated 0.5 mol dm$^{-3}$ H$_2$SO$_4$; loading of Pt: 0.05 mg cm$^{-1}$.

**Figures**

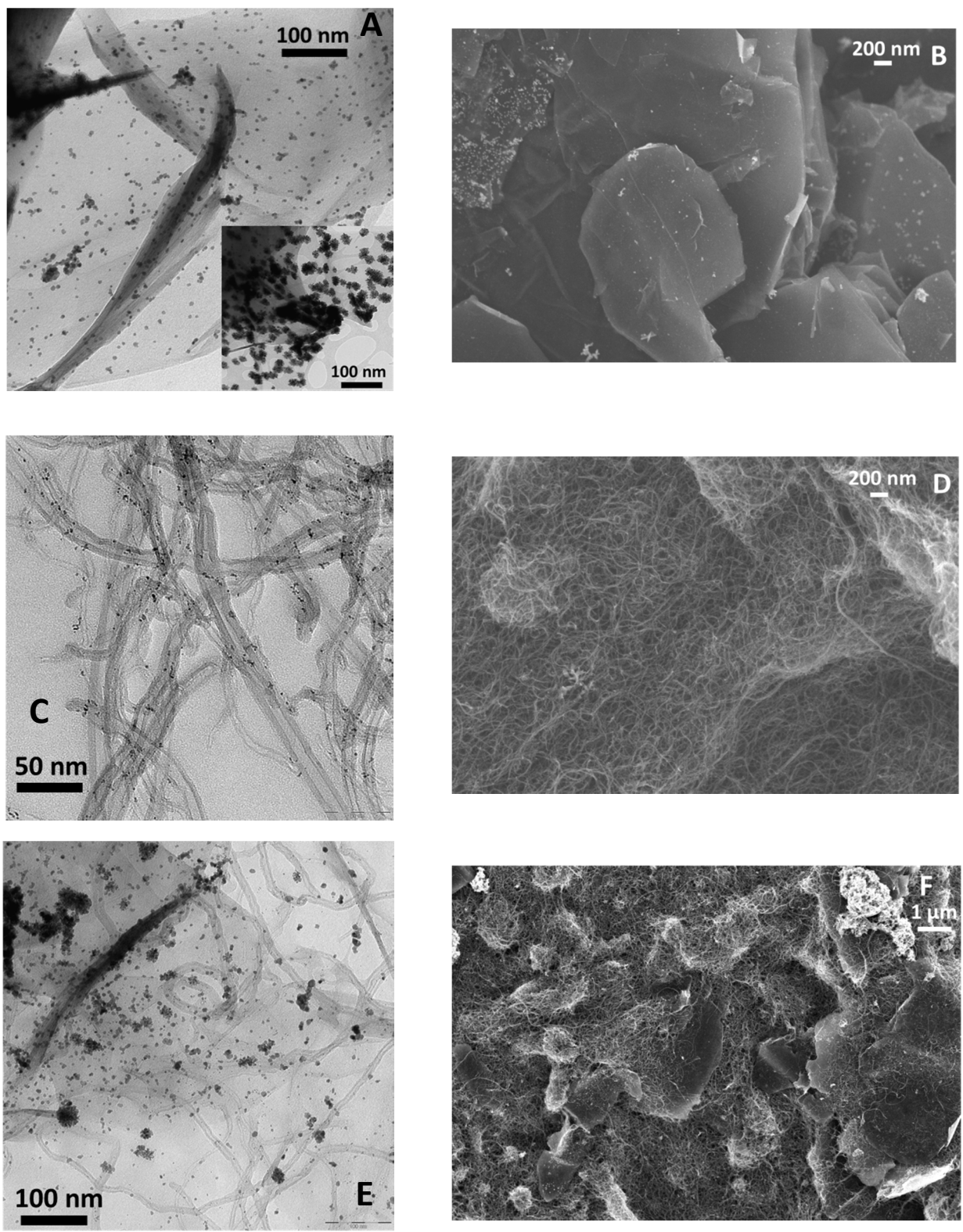

**Fig. 1.** TEM and SEM images of (**A,B**) Pt/rGO,(**C,D**), Ir/CNT and (**E,F**) Pt/rGO+Ir/CNT.

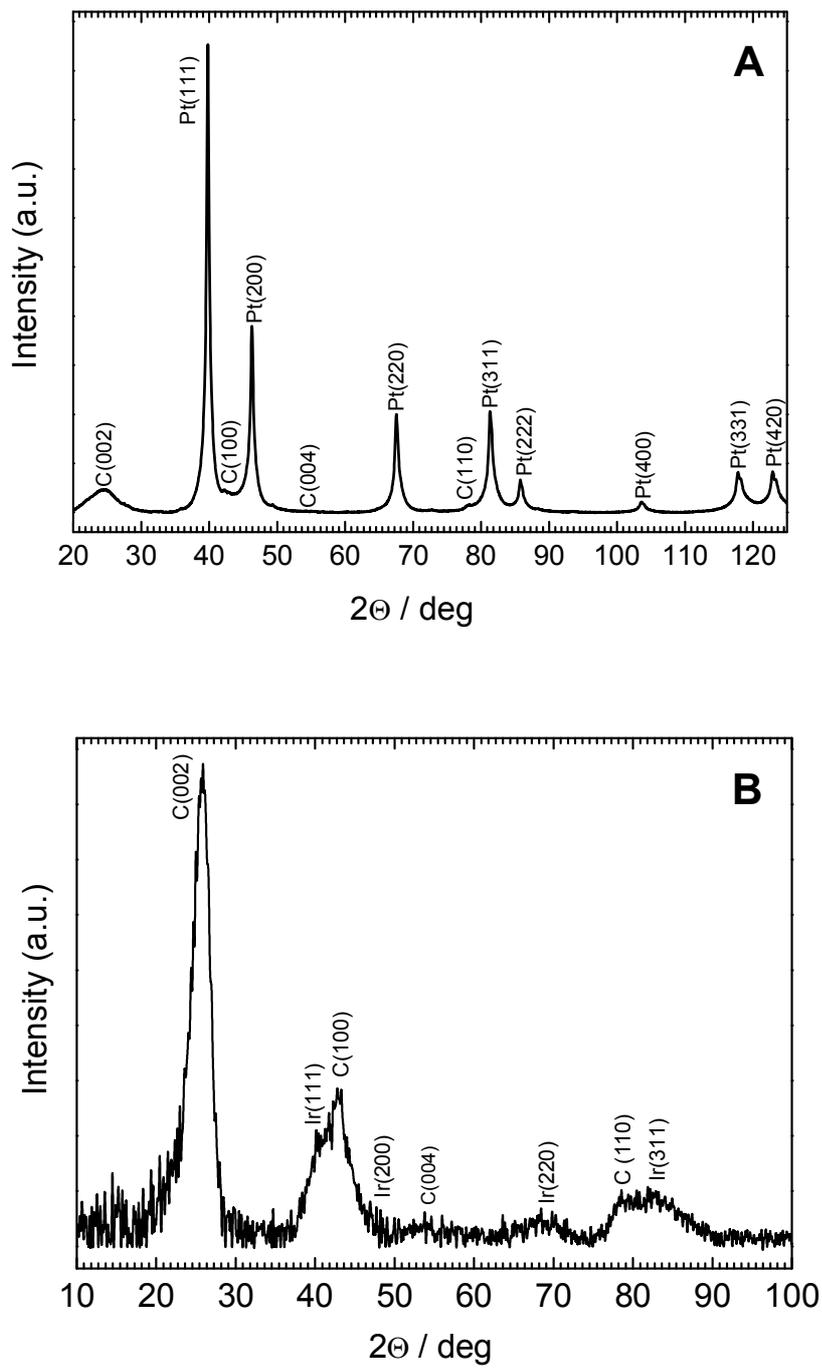

**Fig. 2.** XRD patterns of Pt/rGO(**A**) and Ir/CNT (**B**).

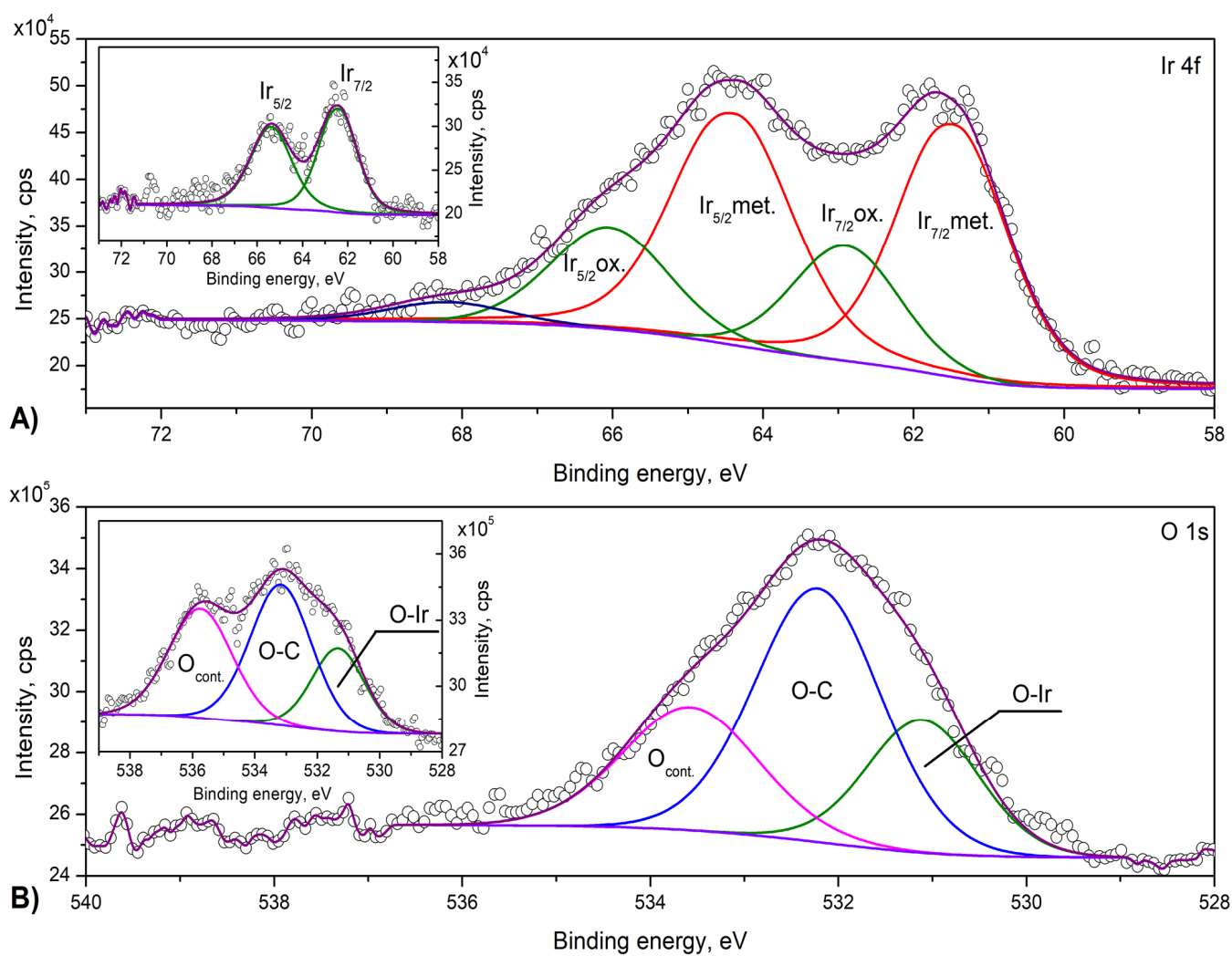

**Fig. 3.** XP spectra of the Ir 4f (**A**) and O 1s (**B**) regions recorded for the Ir/CNT. The insets to panels (**A**) and (**B**) present the Ir 4f and O 1s regions recorded for 2% Ir control sample.

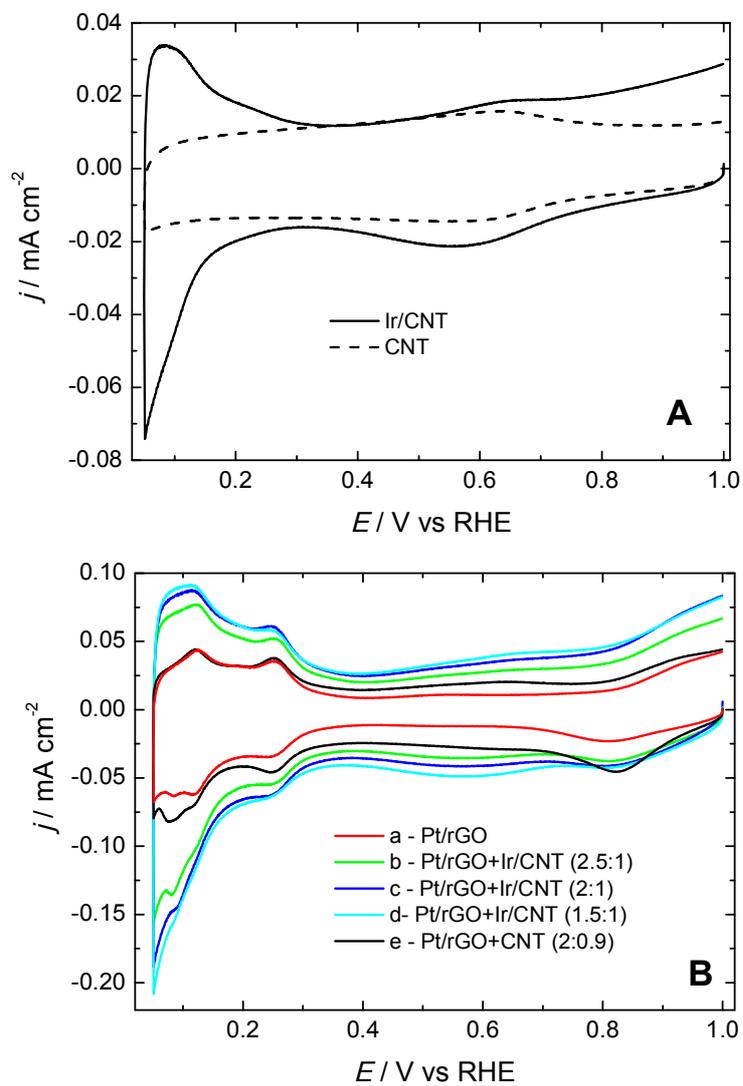

**Fig. 4.** Cyclic voltammetric responses recorded for catalytic films (deposited on glassy carbon disk of RRDE assembly). Electrolyte: argon-saturated 0.5 mol dm$^{-3}$ $H_2SO_4$. Scan rate: 10 mV s$^{-1}$.

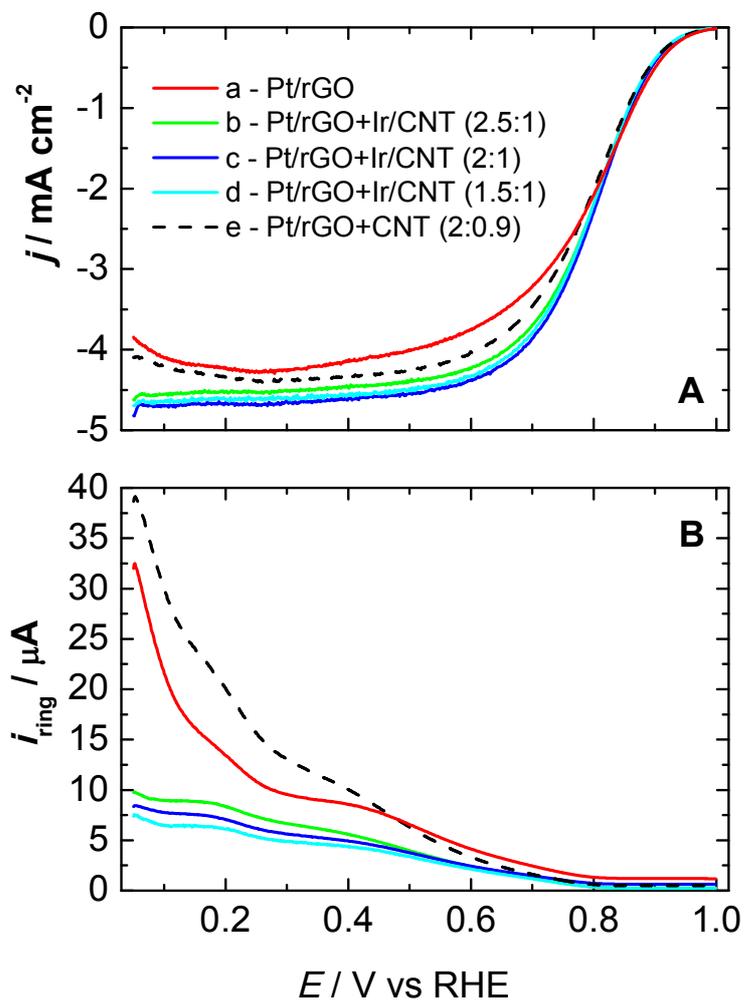

**Fig. 5.** Normalized (background subtracted) rotating disk (**A**) and ring (**B**) voltammograms recorded during oxygen reduction at the catalytic films in oxygen-saturated 0.5 mol dm$^{-3}$ $H_2SO_4$ at the scan rate of 10 mV s$^{-1}$ and rotation rate of 1600 rpm.

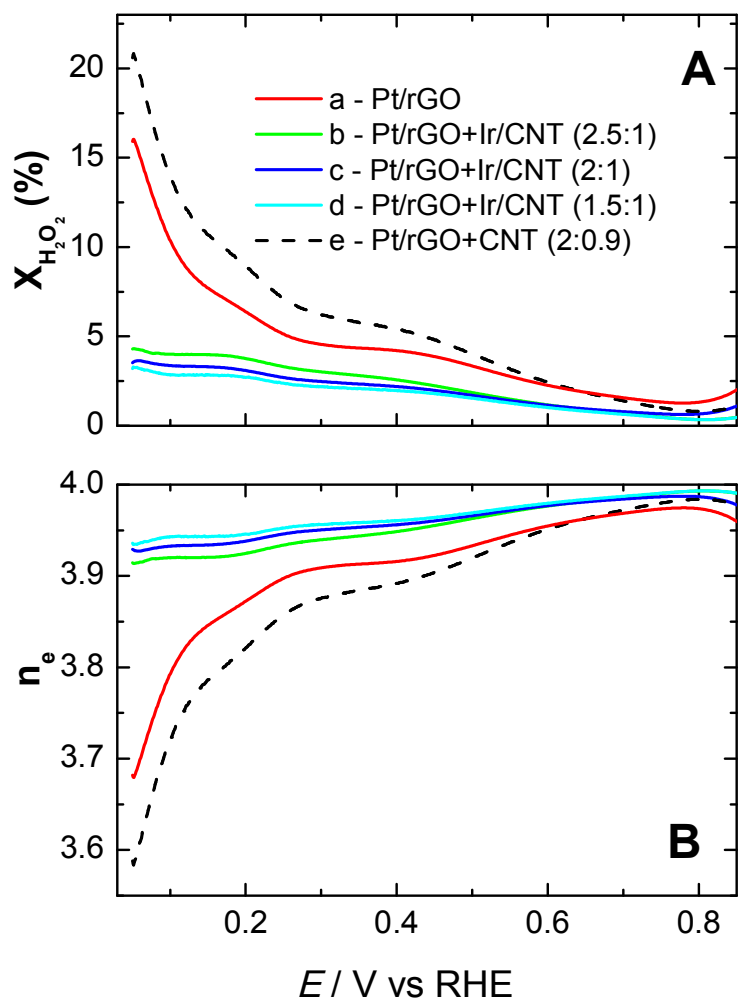

**Fig. 6.** Formation of the hydrogen peroxide intermediate (**A**) and number of exchanged electrons (**B**) during oxygen reduction under conditions of RRDE voltammetric experiments in Fig. 5.

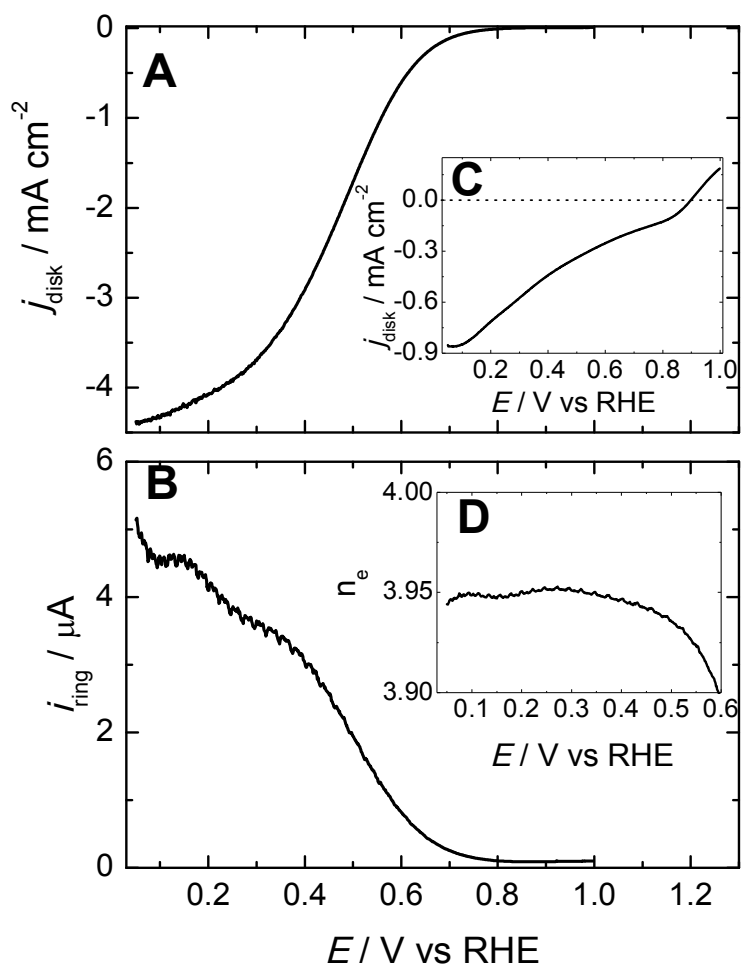

**Fig. 7.** Normalized rotating disk (**A**) and ring (**B**) voltammograms recorded during oxygen reduction at the Ir/CNT film in oxygen-saturated 0.5 mol dm$^{-3}$ H$_2$SO$_4$ at the scan rate of 10 mV s$^{-1}$ and rotation rate of 1600 rpm. (**C**) Normalized rotating disk voltammogram recorded at the Ir/CNT film in 0.5 mol dm$^{-3}$ H$_2$SO$_4$ in the presence of 1 mmol dm$^{-3}$ H$_2$O$_2$; scan rate: 10 mV s$^{-1}$; rotation rate: 1600 rpm. (**D**) The number of exchanged electrons during oxygen reduction at Ir/CNT under conditions of RRDE voltammetric experiment.

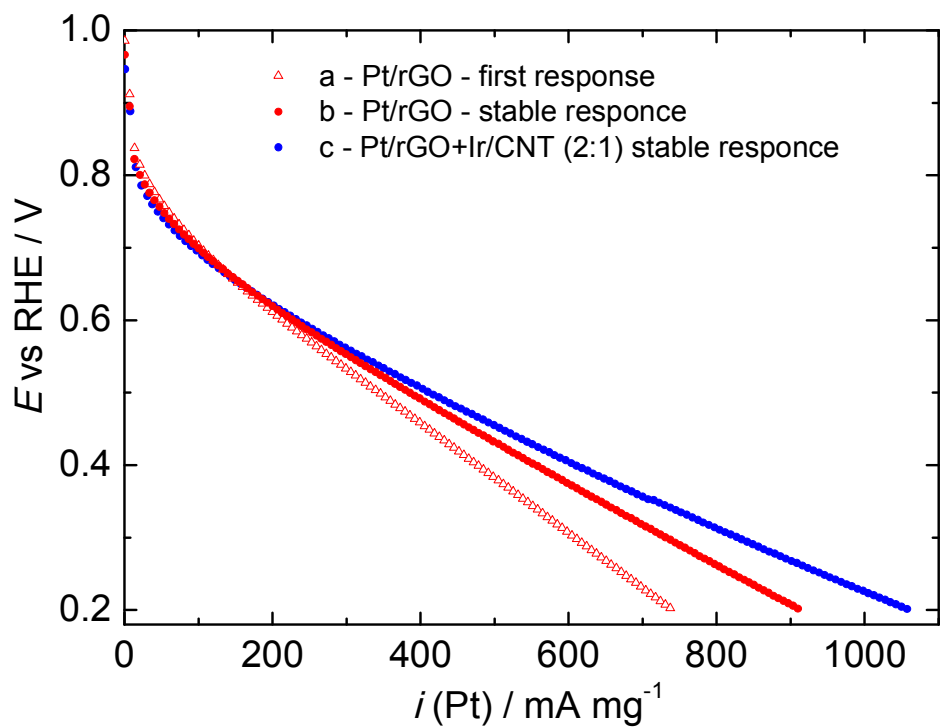

**Fig. 8.** Galvanodynamic steady-state polarization responses for the oxygen reduction on gas diffusion electrodes. Oxygen flux: 50 ml min$^{-1}$, temperature: 80 °C, electrolyte: oxygen saturated 0.5 mol dm$^{-3}$ H$_2$SO$_4$; loading of Pt: 0.05 mg cm$^{-1}$.